\documentclass[pra,showkeys,twocolumn,showpacs]{revtex4-1}
\usepackage{amsmath}
\usepackage{amsfonts}
\usepackage{amssymb}
\usepackage{array}
\usepackage{color}
\usepackage{soul}
\usepackage{graphicx}

\newcommand*{\eq}[1]{(\ref{#1})}

\begin{document}
\title{Solitonic fixed point attractors in the complex Ginzburg-Landau equation for associative memories}

\author{A.~N.~Pyrkov} 
\email{Email address:pyrkov@icp.ac.ru}
\affiliation{Institute of Problems of Chemical
Physics of Russian Academy of Sciences, Acad. Semenov av. 1, Chernogolovka, Moscow
Region, Russia, 142432}

\author{Tim Byrnes}
\affiliation{New York University Shanghai, 1555 Century Ave, Pudong, Shanghai 200122, China}
\affiliation{State Key Laboratory of Precision Spectroscopy, School of Physical and Material Sciences,East China Normal University, Shanghai 200062, China}
\affiliation{NYU-ECNU Institute of Physics at NYU Shanghai, 3663 Zhongshan Road North, Shanghai 200062, China}
\affiliation{National Institute of Informatics, 2-1-2 Hitotsubashi, Chiyoda-ku, Tokyo 101-8430, Japan}
\affiliation{Department of Physics, New York University, New York, NY 10003, USA}

\author{V.~V.~Cherny} 
\affiliation{Institute of Problems of Chemical Physics of Russian Academy of Sciences, Acad. Semenov av. 1, Chernogolovka, Moscow
Region, Russia, 142432 }

\date{\today}

\begin{abstract}
It was recently shown [V.V. Cherny, T. Byrnes, A.N. Pyrkov, \textit{Adv. Quantum Technol.} \textbf{2019} \textit{2}, 1800087] that the nonlinear Schrodinger equation with a simplified dissipative perturbation of special kind features a zero-velocity solitonic solution of non-zero amplitude which can be used in analogy to attractors of Hopfield's associative memory. 
In this work, we consider a more complex dissipative perturbation adding the effect of two-photon absorption and the quintic gain/loss effects that yields formally the complex Ginzburg-Landau equation (CGLE). We construct a perturbation theory for the CGLE with a small dissipative perturbation and define the behavior of the solitonic solutions with parameters of the system and compare the solution with numerical simulations of the CGLE. We show that similarly to the nonlinear Schrodinger equation with  a simplified dissipation term, a zero-velocity solitonic solution of non-zero amplitude appears as an attractor for the CGLE. In this case the amplitude and velocity of the solitonic fixed point attractor does not depend on the quintic gain/loss effects. Furthermore, the effect of two-photon absorption leads to an increase in the strength of the solitonic fixed point attractor.

\end{abstract}

\maketitle

\section{Introduction}

Neuromorphic computing ---  the study of information processing using articficial systems mimicking neuro-biological architectures --- has attracted a huge  amount of interest in modern information science \cite{monroe2014,zhao2010,mead1990,sheridan2017,sebastian2017}. With the recent explosion of interest in quantum information processing systems, it is of great interest whether neuromorphic computing can be combined with quantum approaches \cite{nielsen2000,preskill2018,lamata2019}. One of the best-known model systems in neuromorphic computing is the Hopfield's associative memory \cite{hopfield1982}, which can be considered as a dissipative dynamical system with the ability to make associations \cite{izhikevich2007, strogatz2001, hertz1991}. In this case, the input state is one of stored patterns distorted by noise, and the convergence to the attractor can be understood as recognition of the distorted pattern. Hopfield's associative memory is usually applied to store finite dimensional vectors with dynamics described by a system of ordinary differential equations, which places  restrictions on the patterns that can be processed. It has also been shown that it is possible to encode the information in attractors within an infinite dimensional dynamical system with functional configuration space \cite{src4}.  This allows for the storage and recovery of quite complex and strongly distorted data structures. However, for partial differential equations of a relatively general form, there are no algorithms for the determination of the desired values of system parameters, which turn a given point of functional space into an attractor. 

Meanwhile, it was shown that certain solitonic evolutionary partial differential equations, which admit solutions of the form of localized waves with complex topological structures, can be applied to machine learning \cite{Behera2005}. Since such equations are usually conservative, their solutions necessarily describe the relative motion and interaction of a constant number of solitons, which are determined by initial conditions. The nonlinear Schrodinger equation (NLSE) is one of the best known solitonic equations which has already been applied in very different fields of science. It provides impressively precise description of many physical systems, from vortex filaments to superfluids \cite{Onorato2013,Pitaevskii2003,Falkovich2011}. The Gross-Pitaevskii equation is one particular case of the NLSE and captures many aspects of the time evolution of Bose-Einstein condensates (BECs) \cite{Dalfovo1999, mihalache15}. The use of BECs to solve classical optimization problems \cite{Byrnes2012} and perform quantum algorithms \cite{Byrnes2015, pyrkov2013,pyrkov2014,byrnes12,gross12, pyrkov12, pyrkov19, hecht} have also been investigated. Recently, the nonlinear Schrodinger equation (NLSE), which can be realized in BEC, with a simplified dissipative perturbation which creates a frictional force acting on soliton \cite{kivshar89, malomed05, mihalache17,grelu12} was considered in an application to associative memory and pattern recognition \cite{cherny19}. It was shown that the control of the perturbative term allows one to decrease the velocity of soliton to zero and conserve a positive value of its amplitude. The perturbation makes the zero-velocity solitonic solution of non-zero amplitude into an attractor for all evolution trajectories whose initial conditions are moving solitons.  This paves the way to store information in an infinite dimensional dynamical system using principles which are completely analogous to that of Hopfield's associative memory.

In this paper, we consider the complex Ginzburg-Landau equation (CGLE) \cite{aranson02, morales12} and show that it has similar properties as seen in Ref. \cite{cherny19} that can be exploited towards associative memory and pattern recognition.  The CGLE is of interest since it is a model of experimentally accessible systems such as nonlinear optics, which can form the basis of experimental realization of the general approach.  We construct a perturbation theory for CGLE and compare the solution with numerical simulations.  We show that similarly to the simplified model, a zero-velocity solitonic solution of non-zero amplitude appears in the CGLE, and we investigate the behavior of the solitonic solution on various parameter choices.

\section{The complex Ginzburg-Landau equation}

We consider the CGLE with a dissipative perturbation which creates a frictional force on the soliton. We show that control of the term allows us to decrease the velocity of the soliton to zero and retain some positive value of its amplitude. The existence of such a frictional force would mean that the perturbation turns the resting solitonic solution of certain amplitude into an attractor for all evolution trajectories with initial conditions that are moving solitons.  

The CGLE with a small dissipative perturbation reads 
\begin{equation}
\label{eqn0}
iu_{t} + u_{xx} + 2|u|^{2}u = \epsilon(iAu_{xx} + iBu + iC|u|^{2}u + D |u|^{4}u), 
\end{equation}
where the subscripts denote derivatives with respect to the variable, $\epsilon$ is a small parameter characterizing the perturbation, and $A,B,C,D$ are real positive constants. 
The fundamental monosolitonic solution of LHS of \eq{eqn0} are
\begin{equation}
\label{sol1}
f(x,t) = a \text{sech} (a(x - vt - x_{0})) e^{i\frac{1}{2}vx + (a^2 - \frac{1}{4}v^2)t - i\sigma_{0}}, 
\end{equation}
where $a$ is the soliton amplitude, $v$ is its velocity, and $x_{0}$ with $\sigma_{0}$ are determined by the initial position and phase of the pulse. From \eq{sol1} we can see that the left hand side (LHS) of \eq{eqn0} admits moving and steady state solitonic solutions. The small conservative perturbation of the right hand side (RHS) of \eq{eqn0} makes the soliton oscillate around the minimum of this perturbation potential in a similar way to a classical particle.

%
%

From the results obtained in Ref. \cite{src1}, it is natural to expect that the first term on the RHS of \eqref{eqn0} will create a viscous friction force that will slow down and eventually stop the soliton. However, it also would not be surprising for such a frictional force to make the amplitude decay as well. The second dissipative term is known to increase the soliton amplitude without making changes in its velocity. The third and fourth terms describe the effect of two-photon absorption and the quintic gain/loss effects respectively. 

One possible way to realize the dissipation is to use solitons in a BEC. If the confining potential, which stabilizes a BEC with the attractive interactions against collapse, is made asymmetric such that the atoms can only undergo one-dimensional (1D) motion, it has been predicted to have matter–wave soliton solutions \cite{Garcia1998, Reinhardt1997}. For an atomic BEC, the sign and magnitude of the nonlinearity is determined by the scattering length $\alpha$. The interactions are repulsive for $\alpha > 0$ and attractive for $\alpha < 0$. Dark solitons have been observed in BECs for repulsive interactions \cite{Burger1999, Denschlag2000, Dutton2001}. With the use of Feshbach resonances \cite{Chin2010}, adjusting the atom-atom interaction from repulsive to attractive, bright matter-wave solitons and soliton trains were created in a BEC \cite{Khaykovich2002,Strecker2002}. Furthermore, it was shown that matter-wave bright solitons can form entangled states \cite{Tsarev2018}.


\begin{figure*}[t]
\includegraphics[width=1.8\columnwidth]{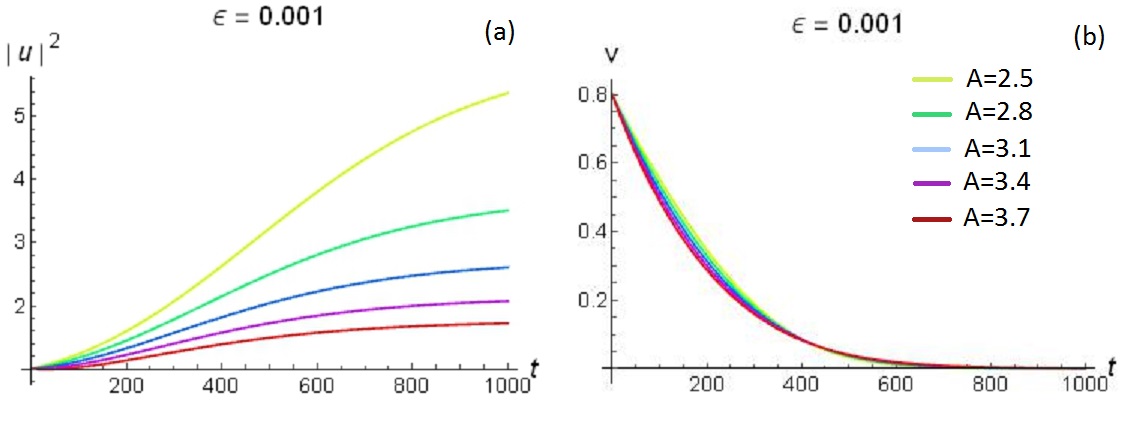}
\caption{(a) Amplitude and (b) velocity of the soliton versus different $A$ for $B=C=1$.} 
\label{diff}
\end{figure*}


In an implementation with a fiber laser the dissipation can be produced in the following way. The combined effects of self-phase modulation and cross-phase modulation induced on two orthogonal polarization components produces a non-linearity during the propagation of the pulse in the fiber. A polarization controller is adjusted at the output of the fiber such that the polarizing isolator passes the central intense part of the pulse but blocks the low-intensity pulse wings. 
In the regime where low-intensity waves are not as efficiently filtered out, the existence of a continuous wave (cw) component that mediates interactions between solitons strongly affect the dynamics and a large number of quasi-cw components produce a noisy background from which dissipative solitons can be formed in the fiber laser cavity and reach the condensed phase. The soliton flow can be adjusted by manual cavity tuning or triggered by the injection of an external low-power cw laser \cite{grelu12,malomed16,mihalache17,belhache03, komarov07, tang01, chouli09, braham17}.

\section{Solitonic fixed point attractors}

We apply Lagrangian perturbation theory for conservative partial differential equations  to describe the soliton behavior under the chosen perturbation. In this case we assume that the solution of the perturbed equation with a single soliton initial condition continues to have this form under evolution but the four characterizing parameters become time-dependent. This assumption is valid for sufficiently small values of $\epsilon$. Thus, we can rewrite this solution in the following form
%
%
\begin{equation}
\label{eqn3}
u(x,t) = a \text{sech} (a\theta) e^{i\xi\theta+i\sigma}  ,
\end{equation}
where $\theta = x - 2\xi t - x_{0}$, $\xi = \frac{v}{2}$, $\sigma = (a^2 + \xi^2)t - \sigma_{0} + \xi x_{0}$.  The LHS of \eq{eqn0} describes a conservative complex scalar field and can thus be (along with its complex conjugate equation) derived from the Lagrangian density
\begin{equation}
\mathfrak{L} = \frac{i}{2}(u^{*}u_{t} - u^{*}_{t}u) - |u_{x}|^{2} + |u|^{4}, \label{l_eqn}
\end{equation}
where $u^{*}$ denotes the complex conjugate to the $u$ field variable. The field $u$ and its complex conjugate $u^{*}$ can be considered independent fields. For this reason they can be taken as the generalized Lagrangian coordinates of our problem. By making a variation of $\mathfrak{L}$ by $u^{*}$ and $u$ we can obtain the corresponding densities of generalized momentum with conventional expressions from Hamiltonian mechanics:
\begin{align}
\pi = \frac{\partial\mathfrak{L}}{\partial u_{t}} = +\frac{i}{2}u^{*}\\
\pi^{*} = \frac{\partial\mathfrak{L}}{\partial u_{t}^{*}} = -\frac{i}{2}u,
\end{align} 
such that the Hamiltonian density takes the form
\begin{equation}
\mathfrak{H} = u_{t}\pi + u^{*}_{t}\pi^{*} - \mathfrak{L} = |u_{x}|^{2} - |u|^{4}.
\end{equation}

The complete energy functional is then given by
\begin{equation}
H = \int_{-\infty}^{\infty}\mathfrak{H}dx .
\end{equation}
Then we can obtain the following equation
\begin{align}
\frac{dH}{dt} & = -\int_{-\infty}^{\infty}(u_{xx}+2|u|^{2}u)u^{*}_tdx + \text{c.c.} \nonumber \\
& = -\epsilon\int_{-\infty}^{\infty}iRu^{*}_{t}dx + \text{c.c.}, \label{en_eqn}
\end{align}
where c.c. denotes the complex conjugate expression and $R$ is a perturbation term of a general form.

To apply perturbation theory to our case, we assume parameters of the soliton to be time-independent, 
and calculate the complete field lagrangian for the single-soliton initial condition under this assumption. This can be done through direct substitution of \eqref{eqn3} into the lanrangian density \eqref{l_eqn} with subsequent integration over the coordinate space:
\begin{align}
L = \int_{-\infty}^{\infty}\mathfrak{L}dx + \text{c.c.} \label{lagr_eqn}
\end{align}
This procedure gives the following expression for the lagrangian in terms of soliton parameters:
\begin{equation}
L = \frac{2}{3}a^{3} - 2a\xi^{2} + 2\alpha\xi\frac{d\alpha}{dt} - 2a\frac{d\sigma}{dt} \label{l_eqn1},
\end{equation}
where $\alpha = x - \theta = 2\xi t + x_{0}$ is the fourth independent parameter of the soliton.
Is is now convenient to rewrite these parameters as a four-dimensional tuple
\begin{equation}
\{a,\xi,\alpha,\sigma\} = \{y_{1}(t),y_{2}(t),y_{3}(t),y_{4}(t)\} .
\end{equation}
%

\begin{figure*}[t]
\includegraphics[width=1.8\columnwidth]{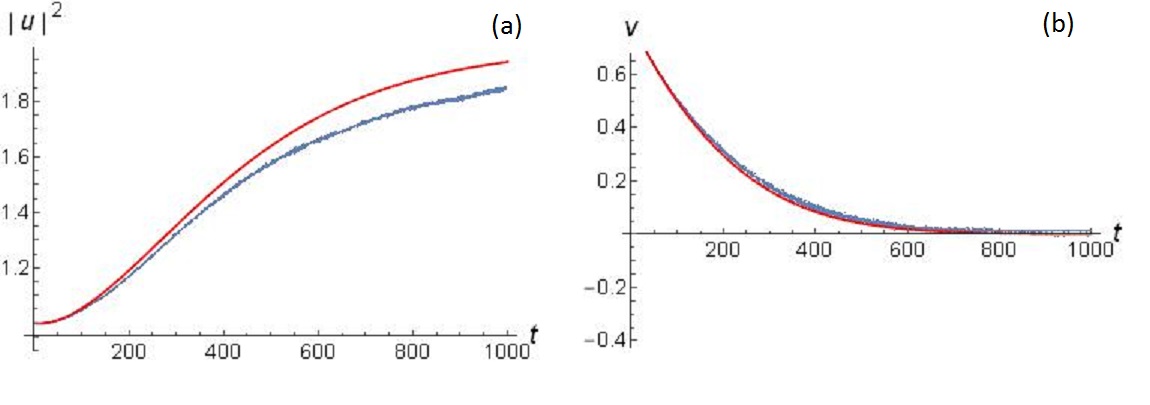}
\caption{Evolution of soliton amplitude (a) and velocity (b) with time. Comparison of perturbation theory (red solid line) with numerical solution (blue dot line) for $\epsilon=0.001$. Parameters of initial soliton are $a = 1, \xi = \frac{4}{5}$ and dissipative parameters are $A=3.5, B=C=1$.} 
\label{exact}
\end{figure*}

In this notation, the main equation in our case has the following form \cite{src2}:
\begin{equation}
\frac{\partial L}{\partial y_{i}} - \frac{d}{dt}(\frac{\partial L}{\partial y_{i,t}}) = \epsilon\int_{-\infty}^{\infty}iR\frac{\partial u^{*}}{\partial y_{i}}dx + \text{c.c.} \label{el_eqn}
\end{equation}
Thus for $R = Au_{xx} + Bu + C|u|^{2}u - i D |u|^{4}u$, we can obtain the system of ordinary differential equations for the parameters of the perturbed soliton
\begin{align}
\dot{\sigma} & = a^{2}(1-\frac{8\epsilon D a^2}{9}) + \xi^{2} \label{sig_eqn} \\
\dot{\alpha} & = 2\xi \\
\dot{a} & = \epsilon(\frac{-2A}{3}a^{3} - 2Aa\xi^{2}  + \frac{4C}{3}a^{3} + 2Ba) \label{a_eqn}\\
\dot{\xi} & = \frac{-4\epsilon Aa^{2}\xi}{3} \label{xi_eqn}.
\end{align}
First of all, we analyze the last equation, which determine the evolution of the velocity $\xi$. Since both $\xi$ and $a$ are always non-negative, the RHS of this equation is negative so the velocity decays with time. Assuming that there exists a stationary point, we can identify the left side of the equation with zero, thus with the condition of $a \neq 0$ we obtain
\begin{equation}
\dot{\xi} = 0 \Rightarrow \frac{-4\epsilon Aa^{2}\xi}{3} = 0 \Rightarrow \xi = 0 . 
\end{equation}
That is, there is only one stationary point of zero velocity as required. 

Next consider \eqref{a_eqn} for  the amplitude $a$.  According to above assumptions, the first two terms on the RHS of this equation decrease the amplitude, while the last two increase it. The attainability of the equilibrium between those forces would mean the existence of an attractive stationary point. We assume that such point exists for $t^{*} \gg 1$,  and the following relation holds: $\dot{a}(t \geq t^{*}) \approx 0$.  It then follows that 
\begin{align}
\dot{a} = 0 & \Rightarrow \frac{-A a^{3}}{3} - Aa\xi^{2}  + \frac{2C a^{3}}{3} + Ba = 0 \nonumber \\
& \Rightarrow a = \sqrt{\frac{3(B - A\xi^{2})}{A-2C}}
\end{align}
and for times large enough $\xi = 0$ the expression for $a^{*}$ is
\begin{equation}
a^{*} = \sqrt{\frac{3B}{A-2C}} . 
\end{equation}
This means, that if $A - 2C > 0$ and $B \neq 0$ our system has an attractor in a form of soliton with positive amplitude $a^{*}$ whose velocity equals to zero. From the expressions we can see that velocity and amplitude of the attractor do not depend on the quintic gain/loss effects. At the same time, increasing of C leads to faster slowing down the soliton and increasing of the attractors amplitude.


Fig. \ref{diff} shows a  numerical evolution of the derived ODE system describing the perturbation theory
approximation for different parameters $A, B, C$. We can see that in this case a standing state soliton in the minimum of potential part of perturbation is translated into a attractor for any monosolitonic initial conditions and that for any given values of $A,B,C$ the ODE system has only one attractor.  Since the expression for $a^{*}$ depends only on the $\frac{B}{A-2C}$, time dependences of soliton amplitude and velocity are presented on Figs. \ref{diff}(a) and \ref{diff}(b) for different choices of parameter $A$ and fixed $B,C$. We see from Fig. \ref{diff}  that increasing $A$ causes the velocity of soliton to decrease faster and that the amplitude of the soliton decreases with increasing $A$. Thus, by controlling the parameters $A,B,C$ we can control the amplitude and velocity of soliton towards an attractor that can be designed to store and restore information. Fig. \ref{exact} shows a simulation of the CGLE in the form \eqref{eqn0} together with numerical evaluation of the derived ODE system describing the perturbation theory approximation. We can see that predictions of perturbation theory are in good agreement with the numerical solution.

\section{Conclusions}

We have investigated the CGLE with a small dissipation term both analytically and numerically to realize solitonic fixed point attractors.   We have shown that in this case a standing state soliton in the minimum of potential part of perturbation is translated into a attractor for any monosolitonic initial conditions. It is shown that the control of dissipative perturbation allows us to handle the attractor of system similarly to Ref. \cite{cherny19}, such that it is possible to store and process information. This approach can be realized with solitons in Bose-Einstein condensates and nonlinear optical systems.

\section*{Acknowledgments}

The work is supported by the RFBR-NSFC collaborative program (Grant No. 18-57-53007). T.B. is supported by the Shanghai Research Challenge Fund; New York University Global Seed Grants for Collaborative Research; National Natural Science Foundation of China (61571301,D1210036A); the NSFC Research Fund for International Young Scientists (11650110425,11850410426); NYU-ECNU Institute of Physics at NYU Shanghai; the Science and Technology Commission of Shanghai Municipality (17ZR1443600); the China Science and Technology Exchange Center (NGA-16-001); and the NSFC-RFBR Collaborative grant (81811530112).

\section*{Author Contributions}

A.N.P and V.V.C performed the calculations. All authors discussed the results and wrote the paper.

\section*{Conflict of interest}

The authors declare no conflict of interest.

\section*{Keywords}

Attractor, complex Ginzburg-Landau equation, soliton, quantum machine learning, associative memory.

\end{document}